\documentclass{article}
\usepackage{spconf,amsmath,graphicx,hyperref}
\usepackage{amssymb}
\usepackage{xcolor}
\usepackage{booktabs}   
\usepackage{multirow}
\usepackage{cleveref} 
\crefname{equation}{}{}

\title{Benchmarking Regularization Methods for 3D Radio Tomographic Imaging}
\name{Wanqin Ma, Yijun Chen, Alikhan
Umirbayev, Jichen Zhang, Ross Murch\thanks{Corresponding author: Wanqin Ma, e-mail:wmaag@connect.ust.hk}}
\address{The Hong Kong
University of Science and Technology}
\begin{document}
%
\maketitle
\begin{abstract}
Integrated Sensing and Communication (ISAC) is an important technology for 6G, enabling wireless systems to perceive their physical environment. Radio Tomographic Imaging (RTI) is a potential technique for device-free sensing in ISAC, reconstructing object locations and shapes from Received Signal Strength (RSS) measurements.  It can complement other techniques, such as radar and LiDAR, by providing a narrowband modality with wavelength resolution. Recently, an extended version of RTI, known as the extended phaseless Rytov approximation (x3DPRA), has been developed to enhance RTI reconstruction quality, estimate material parameters, and extend RTI from two to three dimensions (3D). However, the 3D formulation is even more ill-posed than the 2D form, as the number of measurements is far fewer than the number of unknown voxels. In this work, we systematically evaluate and compare three distinct regularization approaches (Ridge, Total Variation, and Tensor Nuclear) within the x3DPRA framework. We provide a detailed performance analysis by evaluating both reconstruction quality and computational runtime across the three methods. Our findings show that Total Variation regularization provides the best reconstruction quality but not the fastest runtime, providing guidance for developing high-resolution 3D RTI systems for future ISAC applications. 
\end{abstract}
\begin{keywords}
3D Radio Tomography Imaging, Regularization, Integrated Sensing and Communication.
\end{keywords}
\section{Introduction}
\label{sec:I}
Integrated Sensing and Communication (ISAC) is a key technology that is proposed for 6G wireless networks. Its goal is to jointly design communication and sensing functionalities for 6G to enable sensing of the physical environment~\cite{ISAC_1, ISAC_2, ISAC_3, ISAC_4}. Device-free sensing plays an essential role in realizing ISAC by enabling networks to perceive the environment without requiring targets to carry active sensors. Its application includes traffic management, human tracking, smart buildings, and environmental monitoring~\cite{device_f,device_f2}. Radio Tomographic Imaging (RTI) is one potential device-free sensing technique for reconstructing attenuation profiles within a monitored region~\cite{RTI}. It can complement other techniques, such as radar and LiDAR, by providing a narrowband modality with wavelength resolution. Fused or hybrid ISAC techniques that include RTI may therefore be very useful. 

Traditional RTI employs transceivers around the sensing region boundary. These transceivers measure Received Signal Strength (RSS), and when objects or humans enter the region, the resulting signal attenuation across multiple links provides spatial information. By collecting and processing these RSS variations, RTI generates a map that reveals the locations and approximate shapes of objects inside the region. Recent research has reconciled RTI with formal electromagnetic approaches, yielding enhanced methods that improve RTI reconstruction accuracy and functionality. Among these approaches, the extended phaseless Rytov approximation (xPRA) has been shown to provide accurate reconstructions of the shape, location, and attenuation of target objects~\cite{Amar_xPRA, Amar_xRTI}. This represents a significant improvement to RTI, as material classification based on attenuation and permittivity enables the sensing system to distinguish among common indoor objects such as water containers, plastic items, and human tissues, thereby providing richer environmental information.

Despite these important developments, both RTI and xPRA have been limited to two dimensions (2D). 2D reconstruction provides accurate results on planes, but loses critical information about object height and vertical structure. This limitation prevents these methods from being applied to three-dimensional (3D) real-world use cases. Moreover, 3D reconstructions are physically more meaningful because it relates to the real world. In our recent work, we have developed a 3D xPRA framework, denoted x3DPRA, to address this issue by extending previous approaches to volumetric reconstruction to recover complete 3D attenuation distributions within the target region~\cite{x3DPRA}.

One of the key issues with 3D formulations of RTI is that the number of available measurements is far smaller than the number of voxels to be reconstructed due to the additional spatial dimension, as compared to 2D. In 3D, the number of unknown voxels scales cubically, while the number of RSS measurements increases only linearly, increasing the ill-posedness of the inverse problem as compared to 2D. Regularization methods offer a possible solution to such ill-posed challenges by introducing prior information and constraints, enabling the problem to be solved via optimization~\cite{Opt, Opt_2}. Although various optimization methods with different regularization strategies have been proposed in the literature~\cite{RidgeRegressionWiki,3DTV,R_Nuclear,R_2DTV,ADMM,PnP_ADMM} and~\cite{Reg_RTI} have compared different regularization methods for 2D RTI, no comprehensive study systematically evaluates these approaches in the specific context of 3D RTI.

In this work, we benchmark three approaches to regularizing 3D RTI using the x3DPRA formulation. The three approaches are Ridge regularization ($\ell_2$ norm)~\cite{RidgeRegressionWiki}, Total Variation regularization~\cite{3DTV}, and Tensor Nuclear regularization~\cite{R_Nuclear}. 
The contributions of our work are:
\begin{itemize}
    \item We systematically evaluate and compare three regularization approaches
    for solving the inverse scattering problem within the x3DPRA framework.
    \item We provide a detailed performance analysis 
    by evaluating both reconstruction quality and computational runtime across the three methods.
    \item Our findings identify the most effective regularization strategy for 3D RTI, guiding the development of sensing systems suitable for ISAC applications.
\end{itemize}

In the following, Section~\ref{sec: M} introduces the 3D RTI problem, Section~\ref{sec: MM} the regularization methods, and Section~\ref{sec: E} provides experimental results. Conclusions follow in Section~\ref{sec: C}.
\section{FORMULATION}
\label{sec: M}
\begin{figure}[htbp]
    \centering
    \includegraphics[width=0.8\linewidth]{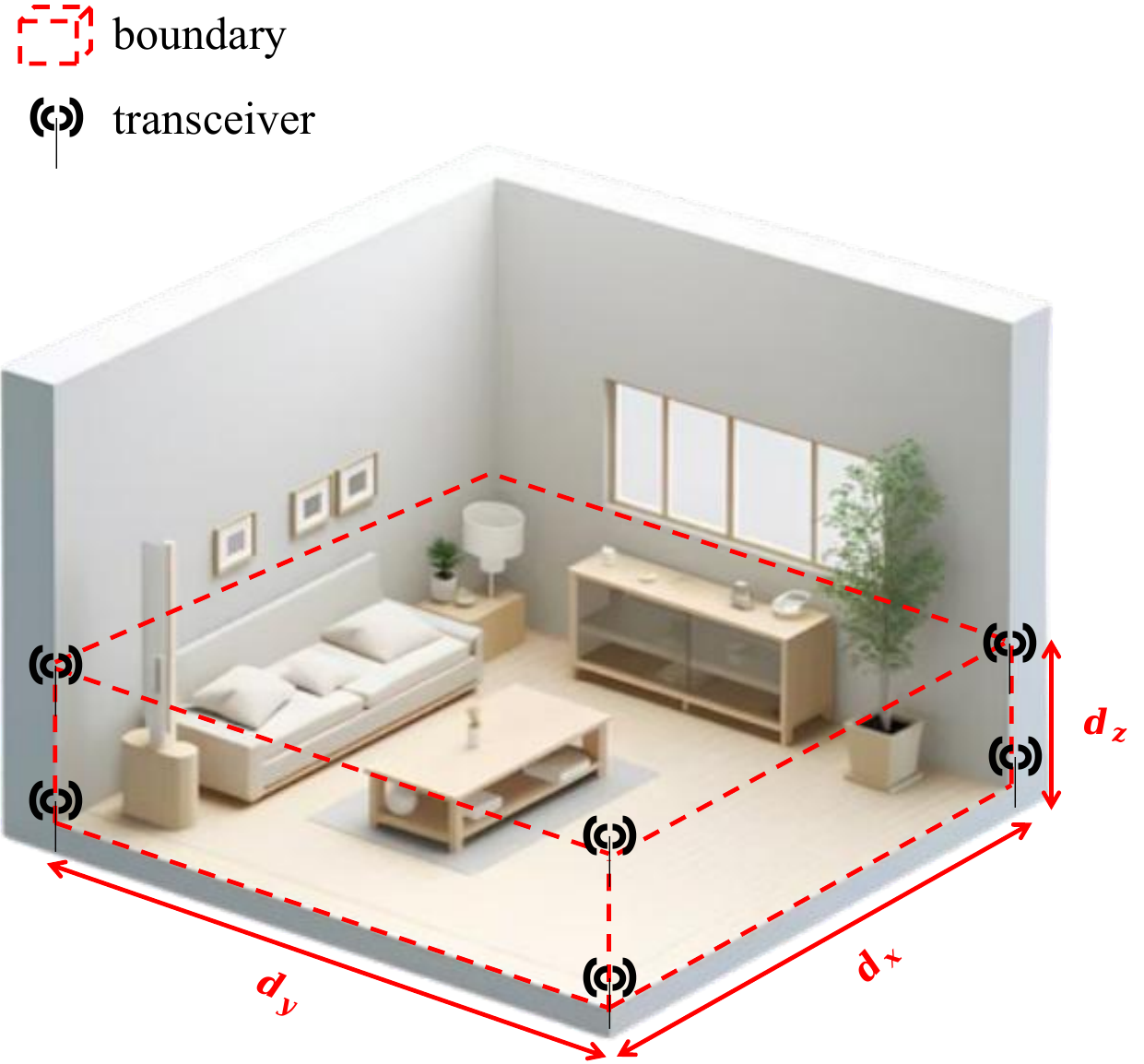}
    \caption{An example indoor scenario with DOI size $d_x \times d_y \times d_z$ and transceivers distributed around the boundary.}
    \label{fig:indoor}
\end{figure}

Consider the indoor scenario in Fig~\ref{fig:indoor} with an example Domain of Interest (DOI) $\mathcal{D} \subset \mathbb{R}^3$ of size $d_x \times d_y \times d_z$. Identical 2.4 GHz Wi-Fi transceivers are deployed around the DOI boundary $\mathcal{B} \subset \mathbb{R}^3$ to collect measurement data. A total of $L$ transceivers are used, creating $M=L(L-1)/2$ communication links (excluding reciprocal links). The DOI is discretized into $N = N_1 \times N_2 \times N_3$ voxels, with size $\Delta v = \Delta d_x \times \Delta d_y \times \Delta d_z$. For $n$th voxel (where $n = 1, 2,\ldots, N$), its relative permittivity is $\epsilon_r(\mathbf{r}_n) = \epsilon_R(\mathbf{r}_n)+j\epsilon_I(\mathbf{r}_n)$. Given that most common objects are low-loss indoor~\cite{Amar_xPRA}, we introduce the attenuation parameter as~\cite{Att_1,Att_2}
\begin{equation}
    \alpha_n = \frac{2\pi\epsilon_I(\mathbf{r}_n)}{\lambda_0 \sqrt{\epsilon_R(\mathbf{r}_n)}} = \frac{2\pi \delta\sqrt{\epsilon_R(\mathbf{r}_n)}}{\lambda_0}
\end{equation}
where the wavelength $\lambda_0 = 0.125$ m is for 2.4 GHz Wi-Fi signals, $\delta = \frac{\epsilon_I(\mathbf{r}_n)}{\epsilon_R(\mathbf{r}_n)}$ represents the loss-tangent.
The formulation, x3DPRA, that we developed~\cite{x3DPRA} provides a straightforward linear model for reconstructing objects in 3D, including shape, location, and attenuation. It is given as
\begin{equation}
\label{eq: sys}
    \mathbf{y} = \mathbf{W}\boldsymbol{\alpha} + \mathbf{n}
\end{equation}
where the received signal strength (RSS) (in dB) for $M$ links is written as $\mathbf{y} = [y_1,  y_2,\ldots, y_M] \in \mathbb{R}^{M\times 1}$, while $\boldsymbol{\alpha}$ denotes the attenuation vector $\boldsymbol{\alpha} = [\alpha_1, \alpha_2, \ldots, \alpha_N] \in \mathbb{R}^{N\times 1}$, and $\mathbf{n} \in \mathbb{R}^{N\times 1}$ is the system noise. $\mathbf{W} \in \mathbb{R}^{M\times N}$ is the x3DPRA model matrix based on electromagnetic theory. For brevity, we omit the explicit form of $\mathbf{W}$; its details can be found in Section III of~\cite {x3DPRA}.

\section{3D REGULARIZATION}
\label{sec: MM}
For inverse scattering problems, such as \cref{eq: sys}, the number of measurements $M$ is typically much smaller than the number of unknowns 
$N$ ($M \ll N$), making the problem ill-posed. To obtain a stable solution, we can formulate it as a regularized least-squares optimization as 
\begin{equation}
    \label{eq: opt}
    \boldsymbol{{\hat \alpha}}=\underset{\boldsymbol{\alpha}}{\operatorname{argmin~}} \frac{1}{2}\|\mathbf{y}-\mathbf{W} \boldsymbol{\alpha}\|_2^2+\gamma R(\boldsymbol{\alpha})
\end{equation}
where $\gamma$ is the regularization parameter and $R(\boldsymbol{\alpha})$ is the regularization term. In this work, we consider three regularization functions for solving~\cref{eq: opt}. 
\subsubsection{3D Ridge Regularization}
Ridge regularization (also known as $\ell_2$ norm) defines the regularization term as $R_{\text{Ridge}}(\mathbf{\boldsymbol{\alpha}}) = ||\mathbf{\boldsymbol{\alpha}}||^2_2$. The solution of $\boldsymbol{\hat{\alpha}}$ can be rewritten as~\cite{RidgeRegressionWiki} 
\begin{equation}
\label{eq: l2}
\boldsymbol{\hat{\alpha}} = \left(\mathbf{W}^{T}\mathbf{W} + \gamma\mathbf{I}\right)^{-1}\mathbf{W}^{T}\mathbf{y}
\end{equation}
where $\mathbf{I}$ is the identity matrix. With prior knowledge of the $\ell_2$-norm, we expect it to produce an overly smooth reconstruction, which cannot preserve sharp edges.

\subsubsection{3D Total Variation Regularization}
Total variation (TV) is commonly used in inverse scattering problems. We follow a recent 3D regularization approach TVReg~\cite{3DTV}. Its regularization term is defined as
\begin{equation}
    \label{eq: TV}
    R_{\text{TV}}(\mathbf{\boldsymbol{\alpha}}) = \sum_{n=1}^N \Phi_\tau\left(D_n \boldsymbol{\alpha}\right)
\end{equation}
where $D_n \in \mathbb{R}^{3 \times N}$ is the discrete difference operator for the differences between voxel $n$ and its neighbors. The Huber function $\Phi_\tau$ is introduced to smooth the problem and is standard in such formulations. The term $D_n\boldsymbol{\alpha}$ is given by
\begin{equation}
        D_n \boldsymbol{\alpha}= \left[\begin{array}{c}\nabla_x \alpha_n \\ \nabla_y \alpha_n \\ \nabla_z \alpha_n\end{array}\right]
\end{equation}
where $\nabla_x \alpha_n = \alpha_{n+1}- \alpha_n$ is the difference along the x dimension and similarly for, $\Delta_y \alpha_n$, and $\nabla_z \alpha_n$. TV has demonstrated strong performance in prior RTI studies~\cite{Amar_xRTI,Reg_RTI}, and we expect it to produce clean reconstructions for our test objects.
\subsubsection{3D Tensor Nuclear Regularization}
The tensor-based nuclear (TN) regularization for multi-dimensional RTI is described in~\cite{R_Nuclear}. To adapt TN for our application, we provide 3D tensor nuclear regularization. The attenuation map $\boldsymbol{\alpha} \in \mathbb{R}^{N\times 1}$, it is reshaped into a tensor $\mathcal{A} \in \mathbb{R}^{N_1 \times N_2 \times N_3}$ as $N = N_1\times N_2 \times N_3$.~\cite{R_Nuclear} defines mode-$n$ unfolding for tensor $\mathcal{A}$ as $\mathcal{A}_{(n)} \in \mathbb{R}^{N_n \times I_n}$, where $I_n=\prod_{i=1, i\neq n}^3 N_i$. By the mode-$n$ matricization, the tensor element $\left(k_1, k_2, k_3\right)$ of $\mathcal{A}$ is mapped to the matrix element $\left(k_n, l_n\right)$ of mode-$n$ unfolding $\mathcal{A}_{(n)}$ by
\begin{equation}
    l_n=\sum_{\substack{i=1 \\ i \neq n}}^3\left(k_i-1\right)\left(\prod_{\substack{j=i+1 \\ j \neq n}}^{4} N_j\right), \quad N_{4}=1
\end{equation}
The regularization penalizes the sum of the nuclear norms of the tensor's mode-$n$ unfoldings.
\begin{equation}
R_{\text{Nuc}}(\boldsymbol{\alpha}) = \sum_{n=1}^{3} \| \mathcal{A}_{(n)} \|_{*}
\end{equation}
This approach encourages a low-rank structure in the data, which can effectively denoise the reconstruction but may also oversmooth the reconstruction.

\section{EXPERIMENTS}
\begin{figure*}[t]
    \centering
    \includegraphics[width=0.8\textwidth,keepaspectratio]{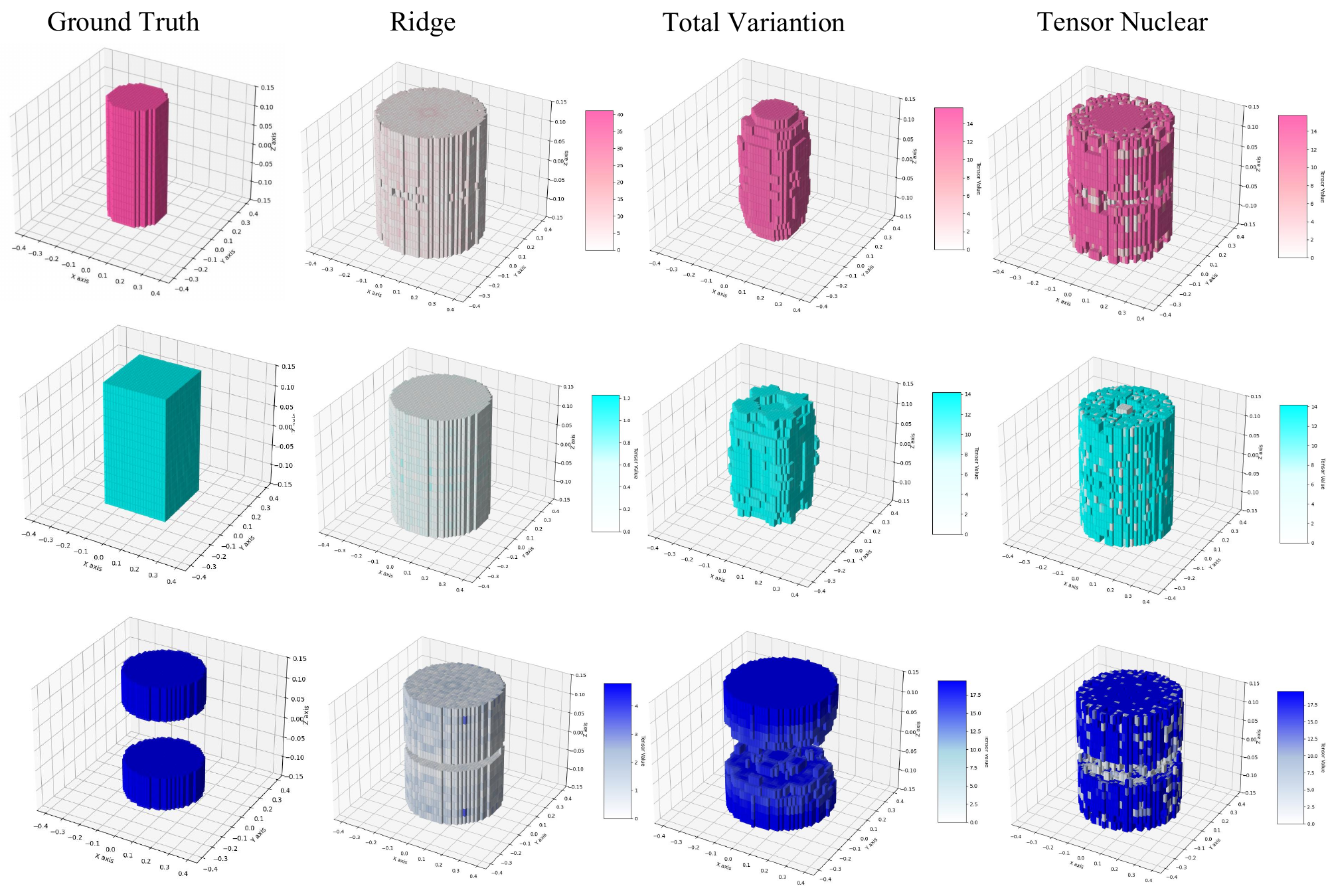}
    \caption{The figure shows the visualization comparison of different optimization methods. The first column is the ground truth. The second, third, and fourth columns show the results from Ridge, TV, and TN, respectively.}
    \label{fig:comp}
\end{figure*}
\label{sec: E}
\subsection{Setup}
We configure a 3D DOI with dimensions $ 0.9 \times 0.9 \times 0.3$ $\text{m}^3$ in CST Studio Suite to simulate RSS measurements. For clarity, we use Cartesian coordinates with the DOI centered at $(0,0,0)$, ranging $x, y \in [-0.45, 0.45]$ m and $z \in [-0.15, 0.15]$ m. The DOI is employed with $L = 48$ dipole antennas, each $6$ cm long, placed at $z = 0, \pm0.15$ m (16 per height, evenly spaced along the boundary). The space between antennas exceeds one wavelength in all simulations to minimize coupling.

In the first column of Fig~\ref{fig:comp}, there are three different objects designed for testing the accuracy and time efficiency of the methods. First, it's a cylinder with a diameter of $0.3$ m and vertical height $0.8$ m. The cylinder is centered within the DOI and extends beyond its vertical boundaries to minimize reflections and scattering from the top and bottom surfaces. Its permittivity is $10+1j$, giving an attenuation parameter $\alpha = 15.8$. The second object is a cuboid with a side length of $0.36$ m, height $0.8$ m, permittivity $8+0.8j$, and attenuation $\alpha=14.2$. The third object consists of two cylinders, each with a diameter of $0.4$ m and a height of $0.25$ m, centered at $(0,0,0.2)$ m and $(0, 0, -0.2)$ m, leaving a $0.15$ m vertical gap to reduce scattering. The object's permittivity is $15+1.5j$, corresponding to $\alpha = 19.5$.

For metrics, we use the Peak Signal-to-Noise Ratio (PSNR) to evaluate reconstruction accuracy and computation time (in seconds) to assess efficiency.
\subsection{Results}
Figure~\ref{fig:comp} shows the reconstructions generated by three methods, and Table~\ref{tab:rti_comparison} lists the metrics. In terms of accuracy, TV achieves the best results among all methods, with an average PSNR of $15.4$ dB. It not only reconstructs the shape close to the ground truth, but also gives accurate attenuation values. However, the results generated by both Ridge and TN are inaccurate. The average PSNR of Ridge is $10.9$ dB, whereas this method mistakenly produces a cylindrical shape in different ground-truth cases and yields incorrect attenuation across all results. Similar results are obtained with TN, which reconstructs the cylinder shape across all settings and yields attenuation values closer to the true value, with an average PSNR of $11.5$ dB. For time efficiency, TN is the fastest method, with an average running time of $8.2$ seconds. Among all methods, Ridge has the highest computational cost, taking around $293.6$ seconds, whereas TV completes in around $18.8$ seconds.
\begin{table*}[t]
\centering
\small  
\caption{Numerical comparison of regularization methods.}
\begin{tabular}{l c c c c c c}
\toprule
\multirow{2}{*}{Object} & \multicolumn{2}{c}{Ridge} & \multicolumn{2}{c}{Total Variation} & \multicolumn{2}{c}{Tensor Nuclear} \\
\cmidrule(lr){2-3} \cmidrule(lr){4-5} \cmidrule(lr){6-7}
 & PSNR (dB) & Time (s) & PSNR (dB) & Time (s) & PSNR (dB) & Time (s) \\
\midrule
Cylinder      & 10.8  & 414.3 & 19.8  & 15.2 & 11.1 & 8.3 \\
Cuboid        & 11.0  & 236.2 & 16.3  & 19.3 & 13.8 & 8.2 \\
2 Cylinders   & 10.9  & 230.4 & 10.0 & 21.8 & 9.58 & 8.1 \\
\bottomrule
\end{tabular}
\label{tab:rti_comparison}
\end{table*}

The three methods perform differently on this task. TV succeeds in balancing accuracy and efficiency because its regularization term $R_{\text{TV}}(\boldsymbol{\alpha})$ aligns with the ground-truth 3D geometries, preserving edges and enabling accurate attenuation reconstruction. It can finish computation relatively quickly because it enforces constancy only within neighboring voxels, saving time on irrelevant variations across voxels. However, it still requires solving 3D subproblems at each iteration by computing the norms of the discrete difference operator ($D_n$), making it slower than TN. Ridge fails because its $\ell_2$ penalty globally shrinks values and blurs boundaries, collapsing distinct shapes into a cylinder. The Hessian $\left(\mathbf{W}^{T}\mathbf{W} + \gamma\mathbf{I}\right)$ remains ill-conditioned in unsolved 3D settings and consumes a lot of computing time. TN fails similarly: Nuclear norms capture the dominant components, so they estimate attenuation levels correctly. However, mode-$n$ unfolding blends distinct features into smooth patterns, failing to preserve 3D boundaries. It helps reduce computational complexity by converting the tensor into matrices, making TN the fastest overall.

In summary, TV is the best approach for reconstructing 3D RTI as it is accurate and fast. Although TN failed on the reconstruction task, it has advantages over TV and Ridge in machine learning (ML) tasks. ML-based RTI first uses simple regularization to produce a blurred reconstruction, then refines it using a network to achieve higher accuracy~\cite{Sami_ML}. In this context, TV yields reconstructions that are overly accurate and preserve loss, making them less suitable as inputs for ML networks. Ridge, on the other hand, is computationally too expensive for generating the large datasets typically required by ML tasks. TN offers blurred results with the lowest computational cost, which suits ML tasks best. Though Ridge is widely used in 2D RTI, it is not suitable for 3D RTI, as it incurs excessive runtime and yields incorrect results.

\section{CONCLUSION}
\label{sec: C}
This work investigated Ridge, TV, and TN regularization methods for solving the 3D RTI problem using our x3DPRA model. TV proves most effective, achieving the highest PSNR in a short runtime. Ridge produces overly smooth reconstructions, while TN also generates over-smooth results, but with the shortest runtime and a good attenuation reconstruction. While TN has computational efficiency, its low-rank constraints blur boundaries, whereas TV preserves correct shapes at a moderate computational cost. Our findings support TV-regularized x3DPRA as an efficient and accurate solution for 3D RTI, while TN has potential for ML-based 3D RTI. Future work will explore integrating machine learning and beamforming with 3D RTI for future ISAC applications.

\newpage
\bibliographystyle{IEEEbib}
\bibliography{strings,refs}

\end{document}